\documentclass[sigconf, screen,nonacm,review=false,timestamp=false]{acmart}

\AtBeginDocument{%
  \providecommand\BibTeX{{%
    \normalfont B\kern-0.5em{\scshape i\kern-0.25em b}\kern-0.8em\TeX}}}

\setcopyright{acmcopyright}
\copyrightyear{2018}
\acmYear{2018}
\acmDOI{10.1145/1122445.1122456}

\acmConference[Woodstock '18]{Woodstock '18: ACM Symposium on Neural
  Gaze Detection}{June 03--05, 2018}{Woodstock, NY}
\acmBooktitle{Woodstock '18: ACM Symposium on Neural Gaze Detection,
  June 03--05, 2018, Woodstock, NY}
\acmPrice{15.00}
\acmISBN{978-1-4503-XXXX-X/18/06}

\usepackage{_macros}
\usepackage{_custom}

\begin{document}

\title[Remote Communication in Distributed Immigrant Families]{Understanding Remote Communication between Grandparents and Grandchildren in Distributed Immigrant Families}

\author{Jiawen Stefanie Zhu}
\email{jiawenz2@uw.edu}
\orcid{0009-0002-2652-7241}
\affiliation{
  \institution{University of Waterloo}
  \city{Waterloo}
  \state{Ontario}
  \country{Canada}
}

\author{Jian Zhao}
\email{jianzhao@uwaterloo.ca}
\orcid{0000-0001-5008-4319}
\affiliation{
  \institution{University of Waterloo}
  \city{Waterloo}
  \state{Ontario}
  \country{Canada}
}

\renewcommand{\shortauthors}{}

\begin{abstract}
Grandparent-grandchild bonds are crucial for both parties. Many immigrant families are geographically dispersed, and the grandparents and grandchildren need to rely on remote communication to maintain their relationships. In addition to geographical separation, grandparents and grandchildren in such families also face language and culture barriers during remote communication. The associated challenges and needs remain understudied as existing research primarily focuses on non-immigrant families or co-located immigrant families. To address this gap, we conducted interviews with six Chinese immigrant families in Canada. Our findings highlight unique challenges faced by immigrant families during remote communication, such as amplified language and cultural barriers due to geographic separation, and provide insights into how technology can better support remote communication. This work offers empirical knowledge about the communication needs of distributed immigrant families and provides directions for future research and design to support grandparent-grandchild remote communication in these families.
\end{abstract}



\begin{teaserfigure}
\centering
    \includegraphics[width = 1\linewidth]{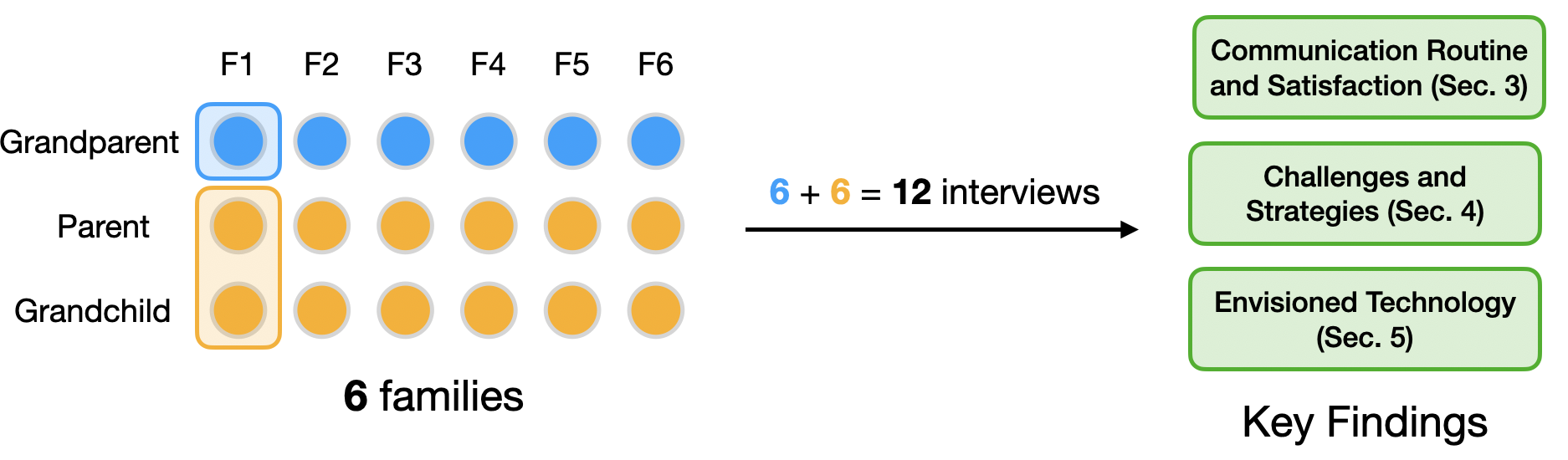}
    \caption{We interviewed six family triads of grandchildren, parent, and grandparent to understand the communication routine and satisfaction, challenges and strategies, and envisioned technology for remote communication between grandparents and grandchildren in distributed immigrant families.}
    \label{fig:teaser}
\end{teaserfigure}

\maketitle

\section{Introduction}

\begin{table*}[tp]
\caption{Participant Demographics, including age and gender of parents (P), grandchildren (GC), and grandparent (GP); self-reported frequency and average duration of remote communication; and estimated language proficiency of grandchildren in Chinese (GC-CN) and grandparents in English (GP-EN) based on interview responses.}
\small
\begin{tabular}{llllllll}
\toprule
\textbf{Family} & \textbf{P} & \textbf{GC} & \textbf{GP} & \textbf{Frequency}      & \textbf{Duration} & \textbf{GC-CN} & \textbf{GP-EN} \\
\toprule
F1              & 40F        & 8M          & 72F         & 1-3 times/week          & 0 - 15 min        & Low            & Low            \\
F2              & 52M        & 11M         & 81M         & 2-3 times/month         & 0 - 15 min        & High           & Low            \\
F3              & 37F        & 7F          & 65F         & 2-3 times/month         & 16 - 30 min       & Medium         & Low            \\
F4              & 46M        & 12M         & 75M         & 4-6 times/week          & 31 - 45 min       & High           & Low            \\
F5              & 51F        & 10F         & 82F         & 2-3 times/month         & 16 - 30 min       & Medium         & Low            \\
F6              & 39F        & 6M          & 67F         & 1+ time/day             & 16 - 30 min       & Low            & Low            \\
\bottomrule
\end{tabular}
\label{tab:participants}
\end{table*}




Grandparent-grandchild bonds hold high emotional importance and provide multi-fold benefits for both parties \cite{familyStorytelling2010}.
As global migration rises \cite{WorldMigrationReport2020}, immigrant families are becoming more common.
Many of these families are geographically dispersed \cite{mazzucatoTransnationalFamiliesWellBeing2011} and the grandparent and grandchildren in these families must rely on remote communication to maintain connections.
This geographical separation is compounded with language and cultural barriers, creating unique challenges that make it more difficult for grandparents and grandchildren in such families to cultivate and maintain their relationships.




Existing work has examined the general problem of remote communication between grandparents and grandchildren in non-immigrant families.
For instance, they have investigated the needs and routines of (non-immigrant) grandparents and grandchildren during remote communication \cite{forghani2014needs}.
Other work focused on activity-based bonding, exploring ways to promote shared experience across distance \cite{stefanidi_magibricks_2023}, support artifact-based storytelling \cite{wallbaum_supporting_2018}, and facilitate flexible micro-sharing \cite{forghani_g2g_2018}.
However, the focus of these work is primarily on overcoming geographical barriers, and do not consider the additional challenges of language and cultural barriers that immigrant families may face.
Another line of research does target grandchildren and grandparents in immigrant families, but it focuses on co-located families \cite{liaqat_hint_2022, liaqatExploringCollaborativeCulture2023a, liaqatParticipatoryDesignIntergenerational2021a}.
In these families, language and cultural barriers are less pronounced compared to in distributed immigrant families, because these families have the opportunity for in-person communication and can naturally build relationships and rapport through everyday interactions.
Additionally, this line of work focuses on conducting shared activities between the grandchildren and grandparents, and does not study regular day-to-day communication.
Thus, the needs and desires of grandparents and grandchildren in distributed immigrant families during everyday remote communication remains understudied. 

In this work, we aim to understand the (1) routines and level of satisfaction, (2) challenges and strategies, and (3) the envisioned role and design of technology during remote communication between grandparents and grandchildren in immigrant families.
To investigate these points, we conducted an interview study with six Chinese immigrant families in Canada as an example, as Canada is an immigration country and China is one of the largest immigrant population in it \cite{governmentofcanadaCensusBriefLinguistic2017, governmentofcanadaImmigrationEthnoculturalDiversity2019}. 
For each family, we separately interviewed the (1) parent and grandchild and (2) grandparent, summing to 12 interviews in total.
Through the interviews, we identified unique challenges faced by immigrant grandparents and grandchildren, including language and cultural barriers, which are further exacerbated by geographical separation.
Additionally, our participants provided insights into the role and design that are desired for technologies for supporting remote communication.

Overall, this paper contributes \textbf{empirical knowledge} about remote communication between immigrant grandchildren and grandparents and provides future \textbf{research and design directions} for supporting their communication.

\section{Methodology}

To understand the patterns, challenges, and strategies of immigrant GP-GC pairs and how they envision technology during remote communication, we conducted an interview study with immigrant families.

\textbf{Participants.}
We recruited six family triads (F\#) of parent (P), grandchild (GC), and grandparent (GP), using mailing lists from a local culture and extracurricular center, and convenience and snowball sampling.
All families were distributed, with the (1) parents and grandchildren and (2) grandparents living separately, for at least 2 years in the past 5 years.
All parents and grandchildren reside in Canada.
Four of the grandparents primarily live in China, one lives 
(separately) in Canada, and one alternates between Canada and China.
The children were all between 8 - 12 years old ($\mu = 9, SD = 2.37$; 2 girls, 4 boys), and born and raised in Canada.
The parents and grandparents grew up in China.
The dominant language of the grandchildren is English, and that of the parents and grandparents is Chinese.
All the parents were fluent in English whereas the grandparents had no working knowledge of English.
Details can be found in Table \ref{tab:participants}.

\textbf{Procedure.}
We conducted semi-structured interviews with the (1) parents and grandchildren and (2) grandparents separately, since given logistics such as timezone differences, it was difficult to interview all three parties together.
We included parents as they are known to take the important role of mediator in the communication of the grandchildren and grandparents \cite{forghani2014needs}.
Each interview lasted roughly 30 - 75 minutes ($\mu = 53.75, SD = 14.94$ min), and was administered remotely via Zoom or WeChat.
Consent (adults) or assent (children) was obtained before the study.
Participants began the study by completing a questionnaire that collected their demographic information and key details about their remote communication habits.
Then we proceeded to the interview, where we asked participants about the (1) communication routine and satisfaction, (2) challenges and strategies, and (3) envisioned role and design of technology, for the remote communication between the grandparents and grandchildren.
The interviews were conducted by the first author in either English, Chinese, or a mix of both, depending on what participants preferred.
The interviews were voice recorded and transcribed verbatim.
Chinese text was machine translated into English and double-checked by the first author.
We open-coded all interview transcripts, since our goal was to gain a systematic and structured understanding \cite{braunUsingThematicAnalysis2006a} of remote communication between immigrant grandchildren and grandparents.
The initial codes were discussed and iteratively refined until all the authors reached an agreement.
The resulting findings are presented in Sections 3 - 5.
This study was approved by our university's ethics board, and participants were remunerated at a rate of \$20 per hour.
\section{Results: Communication Routine and Satisfaction}

\subsection{Structure of Communication}

\textbf{Video calls were the most common and preferred method of communication}, used by all families, as \pqt{[they] can see the person and communicate directly on the spot}{F3, GP}.
Four families (F1, F3, F4, F6) leaned towards regularly scheduling video calls, for example \pqt{trying to call every weekend}{F3, P} or \pqt{chatting almost every dinner}{F6, GP}.
The other two families (F2, F5) preferred initiating conversation ad hoc, \pqt{checking availability when [they] think of it}{F5, P} or \pqt{when [they] haven't talked in a week or so}{F2, P}.

When synchronous communication is not possible, \textbf{messaging remains as a more casual channel} used \pqt{whenever there's something to share}{F3, P}.
Voice, image and video messages were most common, sent as
\pqt{something interesting comes up, such as when the child has something to be proud of [...] for example, winning a swimming competition}{F5, P}
Parents are the mediators in sending these messages, given the young age and limited language skills of the children.
Grandparents were typically on the receiving end for such messages considering their limited technology literacy and
\pqt{there's no expectation for them to respond immediately.}{F3, P}.
F2 added that \pqt{when they [grandparents] want to know more, they'll ask to call}{P}.
Outside of calls and messages, F1 \pqt{maintains an iCloud [cloud service] where we [parents] often upload images and videos [...] it's helpful for the grandparents to keep up with the child's growth}{P}.

\textbf{Chinese is the default language of communication} among immigrant GP-GC pairs, as the the grandchildren's Chinese is generally better than grandparent's English (Table \ref{tab:participants}) and the \pqt{grandparents don't speak English and it's harder for them to learn [English] than for the child to learn Chinese}{F3, P}.
In two families (F1, F6), the grandchild primarily talks in English and the grandparent in Chinese, and the parent translates every sentence for them.
In the other families (F2, F3, F4, F5), the grandchild also uses Chinese, though they still need parents to \pqt{explain or translate words for [them]}{F4, GC}, on a needs-basis.
The grandchild's language choice depends on their Chinese skills.




\subsection{Content of Communication}
The topics of communication between grandchildren and grandparents are usually steered by the grandparents or mediating parents, since \pqt{for children, they don't have the ability to lead the conversation}{F1, P}.
As such, \textbf{conversations mostly centred around the lives of the grandchildren}, such as the grandchildren's \pqt{recent daily life, what they’re up to, what they’re eating, or if anything was happening with relatives or friends lately}{F3, P}.
Sometimes, grandchildren may show concern for the grandparent, such as by \pqt{asking about [their] health and inviting [them] to visit}{F5, GP}.
Families would also send well-wishes and greetings to each others during major events like holidays or birthdays.

\textbf{Communication can be non-verbal} as well, to mimic a shared experience.
For instance, when F6 calls during grandchild's dinner, and \pqt{[the child] is eating something like pork, [the grandparent] might say: `Oh, can I have it?' Then [the child] will put the pork in front of the cell phone and [the grandparent] will pretend that they're eating}{F6, P}.
Grandchildren may also share artifacts of their life, such as \pqt{drawings, crafts or musical pieces they've learned on an instrument, and let them [grandparents] observe}{F3, P}.

We also found that the content often takes a backseat to the \textbf{act of communication itself as a meaningful gesture}.
Daily life can be routine and \pqt{we [the family members] know there’s not always something new or exciting going on}{F5, P}.
The calls and messages are more a way to \pqt{make sure everything is fine, basically just catching up. They [the grandchildren and grandparents] don’t really discuss detailed matters.}{F4, P}.
Partially due to this functional focus, communication between the grandchildren and grandparents tend to be short (Table \ref{tab:participants}).


\subsection{Overall Satisfaction}

\begin{figure}
    \centering
    \includegraphics[width = 0.8\linewidth]{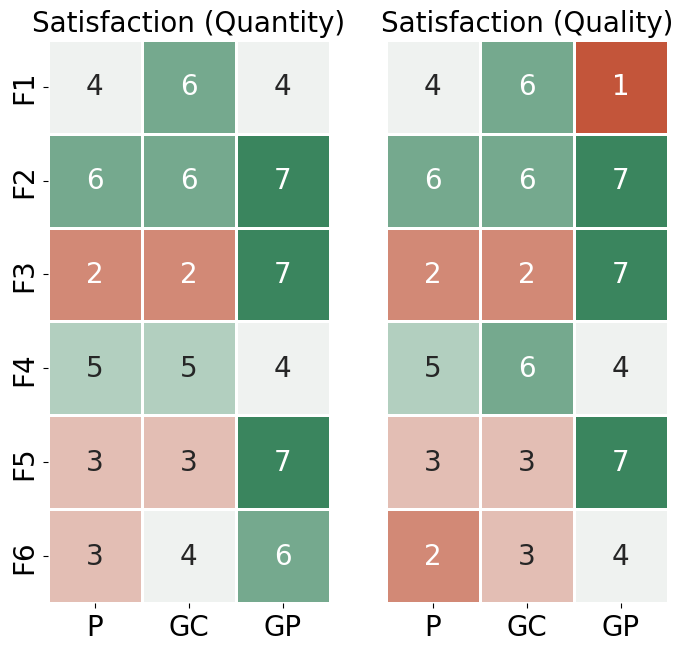}
    \caption{Self-reported satisfaction of the remote communication between grandparents and grandchildren. 1 is the least satisfied and 7 is the most satisfied.}
    \label{fig:sat}
\end{figure}

Satisfaction of the quantity and quality of the remote communication between the grandparents and grandchildren was self-reported individual by the parents (P), grandchildren (GC), and grandparents (GP) on a 7-point Likert scale, where 1 means very dissatisfied and 7 means very satisfied (Figure \ref{fig:sat}).
The average satisfaction was slightly higher for quantity ($\mu = 4.67$) than quality ($\mu = 4.33$).
\textbf{Grandparents were, on average, more satisfied} about the both the quantity and quality of communication ($\mu_{\text{quantity}} = 5.83, \mu_{\text{quality}} = 5.00$) than grandchildren ($\mu_{\text{quantity}} = 4.33, \mu_{\text{quality}} = 4.33$) and parents ($\mu_{\text{quantity}} = 3.83, \mu_{\text{quality}} = 3.66$).

The difference in satisfaction between the grandparents and the grandchildren and parents was likely caused by the Chinese skills of the grandchildren as well as diverging expectations of the different parties.
In families where the satisfaction was on par (F2, F4), the grandchildren had better Chinese skills and were more capable to communicate independently with their grandparents, compared to the other families (F1, F3, F5, F6).

\textbf{Parents and grandchildren tend to have higher expectations of the communication}, worried that the grandparents and grandchildren \pqt{don't really go in depth}{F4, GC} and hoping for \pqt{wider-ranging conversations}{F5, P}.
Parents also expressed long term concerns about the trajectory of the current communication patterns, explaining that \pqt{as they [the grandchildren] grow older, if their language skills aren’t sufficient to communicate, the relationship will gradually become more distant on both sides}{F1, P}.
On the other hand, grandparents tend to have lower expectations of the communication, sharing the sentiment that \pqt{what [they] talk about isn’t important. The key is being able to see them [the grandchildren], joke around a bit, and just enjoy the moment.}{F2, GP}.
The grandparent in F1 is the only one who reported a significantly lower satisfaction than the parents and grandchildren.
They feel discontent that \qt{they're [the grandchild's] there but they don't directly interact with me [...] I have to rely on the parents to translate}, whereas the parents and grandchildren did not perceive the parent scaffolding as significant a problem.


\section{Results: Challenges and Strategies}

\subsection{Language Barriers}

\textbf{Language is the most pronounced challenge }that immigrant grandchildren and grandparents face during remote communication.
While the grandchildren \pqt{can understand Chinese}{F1, P}, but they typically lack the ability to \pqt{expand the conversation or to fully express what they want}{F1, P}.
All six grandchildren faced difficulties in articulating their thoughts and feelings in Chinese, to varying degrees depending on their language skills.

On the one extreme, some grandchildren (F1, F6) \pqt{can say things like hello, but full conversations in Mandarin is not possible}{F1, P}.
In such cases, the parent needs to completely take over the conversation, where \pqt{most of the time I [the parent] asked them [the grandchild] a question in English, they I translate their [the grandchild's] answer to Chinese and the grandparent asked me other questions and I asked him [the grandchild] the questions back}{F6, P}.

More commonly, grandchildren (F2, F3, F4, F5) can convey high-level meanings in Chinese but struggle with \pqt{specific expressions they can't recall}{F3, P}.
For example, the grandchild might try to tell a story about an emergency at school, \pqt{when they want to say that the ambulance is coming, or the police officer shows up, or they want to use a medical term, but they can't find the words in Chinese to allow the grandparents to understand}{F4, P}.
Alternatively, sometimes the grandchild finds a way to express themself, but uses a phrasing that is too unconventional for the grandparent to understand, for example saying \pqt{growing longer instead of growing taller}{F5, P} when talking about their height.

\textbf{Parent scaffolding is the main strategy} used to bridge the language barrier.
Often, parents, knowing both English and Chinese, are asked to provide direct translations or explanations: 
\pqt{I [the grandchild] just always replace words with English words, and I just can't think of a way of fixing that [...] so I would ask my parents to help me translate}{F4, GC}.
However, parents were also concerned about over-reliance, that if they \pqt{jump in too much, they [the grandchild] will lose an opportunity to develop their language skills}{F4, P}.
Thus, another common approach is to guide the grandchild towards \pqt{workarounds that may not be the most concise but still convey the gist}{F3, P}.
For instance, when the grandchild doesn't know the Chinese name of a fruit, \pqt{I [the parent] might ask them to describe what it looks like in Chinese [...] for example if it's red or white, with black spots inside, and what color the skin is. Then, I would say, `that's a dragon fruit,' or `that's a kiwi.'}{F3, P}.

While they address the immediate language barriers, \textbf{these scaffolding strategies do not resolve the motivational challenges} that arise from the considerable effort required for communication.
If the grandparents \pqt{fail to understand for too long, he [the grandchild] would become frustrated and say `forget it, as long as you know it's a thing'}{F4, P}, giving up efforts to communicate completely.
Other children do not communicate with their grandparents when their parent is absent, because \pqt{he [the grandchild] can't reply if he doesn't understand. He just ignores me [the grandparent]}{F6, GP}.
The consensus among the families is that teaching the children Chinese is the best long-term solution, though \pqt{without a language environment, it feels quite difficult}{F5, P}.








\subsection{Cultural Barriers}
\textbf{Culture is also a factor impacting remote communication} between immigrant grandchildren and their grandparents.
Due to growing up and being immersed in different cultures, the grandchildren and grandparents lacked the background to fully understand and connect with each other.
\textbf{This cultural gap limits the depth of their communication and causes problems in initiating and maintaining conversation}.
Families reported that often they \pqt{just can't get on the same page [...] and then it gets a bit awkward because there's no common language. If everyone were on the same page, it would be lively and exciting.}{F5, P}.
For example, when calling during mid-autumn festival, the grandchild struggles to think of anything to say since \pqt{they never encountered something like a rabbit lantern \footnote{Rabbit lantern is a traditional symbol and activity for mid-autumn festival.} [...] it wouldn’t cross their mind to bring up that topic}{F1, P} and the conversation just ends.
As such, many families try to expose their children to their heritage Chinese culture, for example \pqt{telling them more about traditional holidays such as the food and decorations for Spring Festival}{F2, P}.

While they agreed that some level of cultural connection is needed, the \textbf{parents perceived the significance of the cultural gap differently}.
Some parents think lack of cultural connection will hinder the grandchildren and grandparents from fostering a deep and close relationship.
The expression of love is one example, where \pqt{as Chinese people, we usually don't say, I love you. We show it [...] But [the grandchild] didn't understand that's the love Chinese people show [...] He thinks sometimes my mom [the grandparent] is too strict to him}{F6, P}, which negatively influences their dynamics.
Other parents found the cultural gap less important since \pqt{it doesn’t significantly affect [the grandchild's] communication with their grandparents}{F2, P}.
They think that the \pqt{superficial nature of the remote communication between them [the grandparents and grandchildren], despite not being ideal, is the reality}{F4, P}, and that deeper connections, including cultural resonance or sharing life stories, cannot be forced.




\subsection{Geographical Barriers}
While language and cultural barriers can affect in-person communication in immigrant families, \textbf{geographical barriers act as catalysts}, further exacerbating these challenges in remote communication.
Non-verbal communication strategies that were effective for in-person communication are less helpful remotely.
For example, when co-located, grandchildren can express themselves by pointing to things and \pqt{simply saying yes or no, good or not, things like that [...] but conversations like this may not be very desirable or possible in an online setting}{F1, P}.
This highlights the language and cultural barriers, bringing them into sharper focus.

Additionally, \textbf{fostering a meaningful relationship is harder for distributed families} who have to rely on remote communication, because there are less opportunities to engage.
Parents explained that the quality of each individual conversation becomes crucial, because \pqt{if [they] only call once a week and [the grandchild] runs away, that's not very reasonable}{F1, P}.
The grandchildren often had other, more immediate, priorities such as \pqt{wanting to watch TV [...] and might want to just leave}{F5, P}, and the parent had to \pqt{basically force [the grandchild] to sit there, and at least say a couple of words, like `Hello' or something}{F1, P}.
Without strong rapport as a foundation, communication problems are even more difficult to address than they appear at face value.

Ultimately, parents believed that in-person communication cannot be replaced, and \textbf{remote communication is merely a necessary compromise} due to realistic constraints: \pqt{As a family, if I [the parent] want them [the grandparents and grandchildren] to have a deeper connection, I’ll try to plan family trips or activities.} {F4, P}. However, this is often not feasible, and distributed families relying on remote communication face unique challenges as a result.







\section{Results: Envisioned Role and Design of Technology}

\subsection{Address Language Barriers while Maintaining Authenticity}
\textbf{Language barriers were the primary challenges participants hoped technology could help them overcome}, where translation features were widely desired.
In general, participants recognized that \textbf{there was a trade-off between the authenticity of their messages and usage of the translations}.
Some thought that translation tools might deduct from the actual communication experience, as \pqt{they can sometimes get pretty annoying [...] they might distract from me [the grandchild] directly telling them [the grandparent] what I mean}{F4, GC}.
On the other hand, some parents felt that translation features \pqt{would make it easier when [the grandchild] wants to talk to his grandparents}{F2, P}, as the grandchild could \pqt{translate things from English to Chinese when articulating their thoughts, and potentially talk about deeper topics}{F2, P}.
Many grandparents echoed the desire for translation features to facilitate more meaningful exchanges, imagining a scenario where the grandchildren could speak in their dominant language English, while they kept used Chinese.
They decided \pqt{it would be best to be able to communicate fluently. While simple communication is possible now [...] when there's no deep conversation, there might be some problems}{F6, GP}, so \pqt{an app that helps us elderly people [the grandparents] understand the meaning of English, that would be really helpful}{F3, GP}, even if some authenticity might be lost in the translation process.





\subsection{Asymmetrical Design to fit Diverging User Abilities}
\textbf{All families alluded to the differences in technology literacy and learning speed between the grandchildren and grandparents}, and desired technologies to account for these. 
They wanted the technology to be designed so that grandchildren are the ones to actively use its features, while grandparents could remain solely on the receiving end without needing to take any additional actions.
The families pointed out that \pqt{it’s quite difficult for elderly people to learn new technology and [they] can’t keep up with the kids}{F2, GP}.
Additionally, grandparents lack the ability to understand the boundaries of technologies and have difficulties recovering from system errors, illustrating their need for \pqt{the most comfortable, effective, or easier method}{F4, P} possible.
For example, in the case of translation, \pqt{the biggest problem is when I [the grandchild] have to send them [the grandparents] something in English, and they turn it into Chinese [...] I don't think they can figure it out and understand what I mean if the tool makes a mistake. When they send me something in Chinese, and I translate, I feel like I can eventually piece it together}{F2, GC}.
Another example of an asymmetrical design envisioned by participants is a feature for grandchildren to record and retrieve key information from previous conversations, helping to avoid repetitive exchanges that might become frustrating. For instance, \pqt{in a previous conversation, we told the grandparents, `on this date, we’ll have a recital, and [the child] will perform in this program.' Then when they have a similar conversation two weeks later, and the grandparents ask again, this information can be directly shown by the grandchild}{F1, P}.




\subsection{Transitional Technology with Education as Long-Term Goal} 
\textbf{The families saw using technology to overcome their communication barriers as a temporary compromise} to the grandchildren actually learning their heritage language, Chinese.
They wanted technology to be a transitional measure, and contribute towards the longer-term goal of teaching grandchildren the Chinese language.
Parents were worried that over-reliance on translation features could \pqt{gradually weaken his [the child's] initiative to express himself in Chinese}{F3, P}, and stressed that they \pqt{would like to see [the grandchild] has more motivation to talk to his grandparents by himself}{F6, P} in Chinese.
Other \textbf{participants recognized the potential of technology to both bridge communication and provide personalized language learning experiences} for grandchildren.
For example, one parent suggested that \pqt{embedded translation tools could help them [the grandchildren] accumulate vocabulary through practical use. As an simple example, when they [the grandchild] hear the word 'banana' two or three times, they would remember it}{F6, P}. 
They added that \pqt{the ideal result would be that through constant situational learning, [the grandchild] wouldn’t need translation anymore to communicate with their grandparents in Chinese}{F6, P}.

\section{Discussion and Conclusion}

Through an interview study with six distributed immigrant families, we identified the unique patterns, challenges, and needs in remote communication between grandparents and grandchildren in such families, offering insights to guide future research and design.

\subsection{Grounding in Literature: Compare and Contrast to The Non-Immigrant Case}
Comparing our work on remote communication between immigrant grandparents and grandchildren with existing literature on general, non-immigrant families (e.g. \cite{ballagasFamilyCommunicationPhone2009, forghani2014needs, raffleFamilyStoryPlay2010, kennerIntergenerationalLearningChildren2007}), we observed similarities in aspects such as preferred communication mediums, frequency and duration of interactions, and the prevalence of parental scaffolding, likely stemming from shared challenges, such as time zone differences and generational age gaps.
However, we also identified distinct differences between immigrant and non-immigrant cases, highlighting unique challenges faced by immigrant grandparents and grandchildren that warrant further exploration.
Notably, while parental scaffolding is common in both immigrant and non-immigrant families, it is more extensive in immigrant families, extending beyond facilitating conversations to actively participating in them. Additionally, the topics of conversation in distributed immigrant families tend to be more functional, focusing on updating grandparents about the grandchildren, whereas conversations in non-immigrant families seemed topically more balanced \cite{forghani2014needs}. This difference is largely due to the language barriers present in immigrant families, which hinder grandchildren and grandparents from engaging in independent, meaningful conversations.




\subsection{Future Directions}
The insights of this work highlight important future research direction for supporting remote communication between grandchildren and grandparents in immigrant families.
For one, since grandchildren and grandparents typically don't require language support for every utterance, systems supporting their remote communication should explore the concept of minimal language support, offering just enough assistance to sufficiently facilitate the conversation.
The design of such systems need to consider how to preserve authenticity and maintain the grandchild's motivation to express themselves in Chinese, while still providing support for articulation or understanding when necessary.
Next, rather than directly addressing language barriers, future work could focus on providing scaffolding that empowers grandparent-grandchild dyads to independently work toward mutual understanding.
For instance, grandparents could be equipped with guiding questions that help grandchildren articulate their thoughts. This approach could transform the often-frustrating process of overcoming language barriers into an opportunity for shared learning and bonding \cite{kennerIntergenerationalLearningChildren2007}.
Additionally, the history of overcoming language barriers, for example specific vocabulary the grandchild has struggled with in the past, should be viewed as a valuable resource for personalized, in-situ learning. Systems designed to support communication could leverage this history to double as educational tools, thereby enriching learning experiences.



\begin{acks}
This work is supported in part by the University of Waterloo through the MURA program. We thank all the participants for their time and our reviewers for their valuable insights.
We acknowledge that much of our work takes place on the traditional territory of the Neutral, Anishinaabeg, and Haudenosaunee peoples. Our main campus is situated on the Haldimand Tract, the land granted to the Six Nations that includes six miles on each side of the Grand River.    
\end{acks}

\section*{Dedication}

This work is dedicated to grandparents around the world –
\begin{CJK*}{UTF8}{gbsn}姥姥\end{CJK*} (laolao),
\begin{CJK*}{UTF8}{gbsn}姥爷\end{CJK*} (laoye),
\begin{CJK*}{UTF8}{gbsn}爷爷\end{CJK*} (yeye), and
\begin{CJK*}{UTF8}{gbsn}奶奶\end{CJK*} (nainai).

\bibliographystyle{ACM-Reference-Format}
\bibliography{_main.bib}


\begin{thebibliography}{16}


\ifx \showCODEN    \undefined \def \showCODEN     #1{\unskip}     \fi
\ifx \showDOI      \undefined \def \showDOI       #1{#1}\fi
\ifx \showISBNx    \undefined \def \showISBNx     #1{\unskip}     \fi
\ifx \showISBNxiii \undefined \def \showISBNxiii  #1{\unskip}     \fi
\ifx \showISSN     \undefined \def \showISSN      #1{\unskip}     \fi
\ifx \showLCCN     \undefined \def \showLCCN      #1{\unskip}     \fi
\ifx \shownote     \undefined \def \shownote      #1{#1}          \fi
\ifx \showarticletitle \undefined \def \showarticletitle #1{#1}   \fi
\ifx \showURL      \undefined \def \showURL       {\relax}        \fi
\providecommand\bibfield[2]{#2}
\providecommand\bibinfo[2]{#2}
\providecommand\natexlab[1]{#1}
\providecommand\showeprint[2][]{arXiv:#2}

\bibitem[Wor(2041)]%
        {WorldMigrationReport2020}
 \bibinfo{year}{Thu, 05/21/2020 - 20:41}\natexlab{}.
\newblock \showarticletitle{World {{Migration Report}} 2024}.
\newblock  (\bibinfo{year}{Thu, 05/21/2020 - 20:41}).
\newblock
\showISSN{1561-5502}


\bibitem[Ballagas et~al\mbox{.}(2009)]%
        {ballagasFamilyCommunicationPhone2009}
\bibfield{author}{\bibinfo{person}{Rafael Ballagas}, \bibinfo{person}{Joseph~'Jofish' Kaye}, \bibinfo{person}{Morgan Ames}, \bibinfo{person}{Janet Go}, {and} \bibinfo{person}{Hayes Raffle}.} \bibinfo{year}{2009}\natexlab{}.
\newblock \showarticletitle{Family Communication: Phone Conversations with Children}. In \bibinfo{booktitle}{\emph{Proceedings of the 8th {{International Conference}} on {{Interaction Design}} and {{Children}}}} \emph{(\bibinfo{series}{{{IDC}} '09})}. \bibinfo{publisher}{{Association for Computing Machinery}}, \bibinfo{address}{{New York, NY, USA}}, \bibinfo{pages}{321--324}.
\newblock
\showISBNx{978-1-60558-395-2}
\urldef\tempurl%
\url{https://doi.org/10.1145/1551788.1551874}
\showDOI{\tempurl}


\bibitem[Braun and Clarke(2006)]%
        {braunUsingThematicAnalysis2006a}
\bibfield{author}{\bibinfo{person}{Virginia Braun} {and} \bibinfo{person}{Victoria Clarke}.} \bibinfo{year}{2006}\natexlab{}.
\newblock \showarticletitle{Using Thematic Analysis in Psychology}.
\newblock \bibinfo{journal}{\emph{Qualitative Research in Psychology}} \bibinfo{volume}{3}, \bibinfo{number}{2} (\bibinfo{date}{Jan.} \bibinfo{year}{2006}), \bibinfo{pages}{77--101}.
\newblock
\showISSN{1478-0887}
\urldef\tempurl%
\url{https://doi.org/10.1191/1478088706qp063oa}
\showDOI{\tempurl}


\bibitem[Forghani and Neustaedter(2014)]%
        {forghani2014needs}
\bibfield{author}{\bibinfo{person}{Azadeh Forghani} {and} \bibinfo{person}{Carman Neustaedter}.} \bibinfo{year}{2014}\natexlab{}.
\newblock \showarticletitle{The Routines and Needs of Grandparents and Parents for Grandparent-Grandchild Conversations over Distance}. In \bibinfo{booktitle}{\emph{Proceedings of the {{SIGCHI Conference}} on {{Human Factors}} in {{Computing Systems}}}} \emph{(\bibinfo{series}{{{CHI}} '14})}. \bibinfo{publisher}{{Association for Computing Machinery}}, \bibinfo{address}{{New York, NY, USA}}, \bibinfo{pages}{4177--4186}.
\newblock
\showISBNx{978-1-4503-2473-1}
\urldef\tempurl%
\url{https://doi.org/10.1145/2556288.2557255}
\showDOI{\tempurl}


\bibitem[Forghani et~al\mbox{.}(2018)]%
        {forghani_g2g_2018}
\bibfield{author}{\bibinfo{person}{Azadeh Forghani}, \bibinfo{person}{Carman Neustaedter}, \bibinfo{person}{Manh~C. Vu}, \bibinfo{person}{Tejinder~K. Judge}, {and} \bibinfo{person}{Alissa~N. Antle}.} \bibinfo{year}{2018}\natexlab{}.
\newblock \showarticletitle{{G2G}: {The} {Design} and {Evaluation} of a {Shared} {Calendar} and {Messaging} {System} for {Grandparents} and {Grandchildren}}. In \bibinfo{booktitle}{\emph{Proceedings of the 2018 {CHI} {Conference} on {Human} {Factors} in {Computing} {Systems}}}. \bibinfo{publisher}{ACM}, \bibinfo{address}{Montreal QC Canada}, \bibinfo{pages}{1--12}.
\newblock
\showISBNx{978-1-4503-5620-6}
\urldef\tempurl%
\url{https://doi.org/10.1145/3173574.3173729}
\showDOI{\tempurl}


\bibitem[{Government of Canada}(2017)]%
        {governmentofcanadaCensusBriefLinguistic2017}
\bibfield{author}{\bibinfo{person}{Statistics~Canada {Government of Canada}}.} \bibinfo{year}{2017}\natexlab{}.
\newblock \bibinfo{title}{Census in {{Brief}}: {{Linguistic}} Integration of Immigrants and Official Language Populations in {{Canada}}}.
\newblock \bibinfo{howpublished}{https://www12.statcan.gc.ca/census-recensement/2016/as-sa/98-200-x/2016017/98-200-x2016017-eng.cfm}.
\newblock


\bibitem[{Government of Canada}(2019)]%
        {governmentofcanadaImmigrationEthnoculturalDiversity2019}
\bibfield{author}{\bibinfo{person}{Statistics~Canada {Government of Canada}}.} \bibinfo{year}{2019}\natexlab{}.
\newblock \bibinfo{title}{Immigration and Ethnocultural Diversity Statistics}.
\newblock \bibinfo{howpublished}{https://www.statcan.gc.ca/en/subjects-start/immigration\_and\_ethnocultural\_diversity}.
\newblock


\bibitem[Kenner et~al\mbox{.}(2007)]%
        {kennerIntergenerationalLearningChildren2007}
\bibfield{author}{\bibinfo{person}{Charmian Kenner}, \bibinfo{person}{Mahera Ruby}, \bibinfo{person}{John Jessel}, \bibinfo{person}{Eve Gregory}, {and} \bibinfo{person}{Tahera Arju}.} \bibinfo{year}{2007}\natexlab{}.
\newblock \showarticletitle{Intergenerational Learning between Children and Grandparents in East {{London}}}.
\newblock \bibinfo{journal}{\emph{Journal of Early Childhood Research}} \bibinfo{volume}{5}, \bibinfo{number}{3} (\bibinfo{date}{Oct.} \bibinfo{year}{2007}), \bibinfo{pages}{219--243}.
\newblock
\showISSN{1476-718X}
\urldef\tempurl%
\url{https://doi.org/10.1177/1476718X07080471}
\showDOI{\tempurl}


\bibitem[Liaqat et~al\mbox{.}(2021)]%
        {liaqatParticipatoryDesignIntergenerational2021a}
\bibfield{author}{\bibinfo{person}{Amna Liaqat}, \bibinfo{person}{Benett Axtell}, {and} \bibinfo{person}{Cosmin Munteanu}.} \bibinfo{year}{2021}\natexlab{}.
\newblock \showarticletitle{Participatory {{Design}} for {{Intergenerational Culture Exchange}} in {{Immigrant Families}}: {{How Collaborative Narration}} and {{Creation Fosters Democratic Engagement}}}.
\newblock \bibinfo{journal}{\emph{Proceedings of the ACM on Human-Computer Interaction}} \bibinfo{volume}{5}, \bibinfo{number}{CSCW1} (\bibinfo{date}{April} \bibinfo{year}{2021}), \bibinfo{pages}{1--40}.
\newblock
\showISSN{2573-0142}
\urldef\tempurl%
\url{https://doi.org/10.1145/3449172}
\showDOI{\tempurl}


\bibitem[Liaqat et~al\mbox{.}(2022)]%
        {liaqat_hint_2022}
\bibfield{author}{\bibinfo{person}{Amna Liaqat}, \bibinfo{person}{Benett Axtell}, {and} \bibinfo{person}{Cosmin Munteanu}.} \bibinfo{year}{2022}\natexlab{}.
\newblock \showarticletitle{"{With} a hint she will remember": {Collaborative} {Storytelling} and {Culture} {Sharing} between {Immigrant} {Grandparents} and {Grandchildren} {Via} {Magic} {Thing} {Designs}}.
\newblock \bibinfo{journal}{\emph{Proceedings of the ACM on Human-Computer Interaction}} \bibinfo{volume}{6}, \bibinfo{number}{CSCW2} (\bibinfo{date}{Nov.} \bibinfo{year}{2022}), \bibinfo{pages}{268:1--268:37}.
\newblock
\urldef\tempurl%
\url{https://doi.org/10.1145/3555158}
\showDOI{\tempurl}


\bibitem[Liaqat et~al\mbox{.}(2023)]%
        {liaqatExploringCollaborativeCulture2023a}
\bibfield{author}{\bibinfo{person}{Amna Liaqat}, \bibinfo{person}{Carrie Demmans~Epp}, \bibinfo{person}{Minghao Cai}, {and} \bibinfo{person}{Cosmin Munteanu}.} \bibinfo{year}{2023}\natexlab{}.
\newblock \showarticletitle{Exploring {{Collaborative Culture Sharing Dynamics}} in {{Immigrant Families}} through {{Digital Crafting}} and {{Storytelling}}}.
\newblock \bibinfo{journal}{\emph{Proceedings of the ACM on Human-Computer Interaction}} \bibinfo{volume}{7}, \bibinfo{number}{CSCW2} (\bibinfo{date}{Sept.} \bibinfo{year}{2023}), \bibinfo{pages}{1--29}.
\newblock
\showISSN{2573-0142}
\urldef\tempurl%
\url{https://doi.org/10.1145/3610098}
\showDOI{\tempurl}


\bibitem[Mazzucato and Schans(2011)]%
        {mazzucatoTransnationalFamiliesWellBeing2011}
\bibfield{author}{\bibinfo{person}{Valentina Mazzucato} {and} \bibinfo{person}{Djamila Schans}.} \bibinfo{year}{2011}\natexlab{}.
\newblock \showarticletitle{Transnational {{Families}} and the {{Well-Being}} of {{Children}}: {{Conceptual}} and {{Methodological Challenges}}}.
\newblock \bibinfo{journal}{\emph{Journal of Marriage and the Family}} \bibinfo{volume}{73}, \bibinfo{number}{4} (\bibinfo{date}{Aug.} \bibinfo{year}{2011}), \bibinfo{pages}{704}.
\newblock
\urldef\tempurl%
\url{https://doi.org/10.1111/j.1741-3737.2011.00840.x}
\showDOI{\tempurl}


\bibitem[Raffle et~al\mbox{.}(2010)]%
        {raffleFamilyStoryPlay2010}
\bibfield{author}{\bibinfo{person}{Hayes Raffle}, \bibinfo{person}{Rafael Ballagas}, \bibinfo{person}{Glenda Revelle}, \bibinfo{person}{Hiroshi Horii}, \bibinfo{person}{Sean Follmer}, \bibinfo{person}{Janet Go}, \bibinfo{person}{Emily Reardon}, \bibinfo{person}{Koichi Mori}, \bibinfo{person}{Joseph Kaye}, {and} \bibinfo{person}{Mirjana Spasojevic}.} \bibinfo{year}{2010}\natexlab{}.
\newblock \showarticletitle{Family Story Play: Reading with Young Children (and Elmo) over a Distance}. In \bibinfo{booktitle}{\emph{Proceedings of the {{SIGCHI Conference}} on {{Human Factors}} in {{Computing Systems}}}} \emph{(\bibinfo{series}{{{CHI}} '10})}. \bibinfo{publisher}{{Association for Computing Machinery}}, \bibinfo{address}{{New York, NY, USA}}, \bibinfo{pages}{1583--1592}.
\newblock
\showISBNx{978-1-60558-929-9}
\urldef\tempurl%
\url{https://doi.org/10.1145/1753326.1753563}
\showDOI{\tempurl}


\bibitem[Stefanidi et~al\mbox{.}(2023)]%
        {stefanidi_magibricks_2023}
\bibfield{author}{\bibinfo{person}{Evropi Stefanidi}, \bibinfo{person}{Julia Dominiak}, \bibinfo{person}{Marit Bentvelzen}, \bibinfo{person}{Paweł~W. Woźniak}, \bibinfo{person}{Johannes Schöning}, \bibinfo{person}{Yvonne Rogers}, {and} \bibinfo{person}{Jasmin Niess}.} \bibinfo{year}{2023}\natexlab{}.
\newblock \showarticletitle{{MagiBricks}: {Fostering} {Intergenerational} {Connectedness} in {Distributed} {Play} with {Smart} {Toy} {Bricks}}. In \bibinfo{booktitle}{\emph{Proceedings of the 22nd {Annual} {ACM} {Interaction} {Design} and {Children} {Conference}}} \emph{(\bibinfo{series}{{IDC} '23})}. \bibinfo{publisher}{Association for Computing Machinery}, \bibinfo{address}{New York, NY, USA}, \bibinfo{pages}{239--252}.
\newblock
\showISBNx{9798400701313}
\urldef\tempurl%
\url{https://doi.org/10.1145/3585088.3589390}
\showDOI{\tempurl}


\bibitem[Vutborg et~al\mbox{.}(2010)]%
        {familyStorytelling2010}
\bibfield{author}{\bibinfo{person}{René Vutborg}, \bibinfo{person}{Jesper Kjeldskov}, \bibinfo{person}{Sonja Pedell}, {and} \bibinfo{person}{Frank Vetere}.} \bibinfo{year}{2010}\natexlab{}.
\newblock \showarticletitle{Family storytelling for grandparents and grandchildren living apart}. In \bibinfo{booktitle}{\emph{Proceedings of the 6th {Nordic} {Conference} on {Human}-{Computer} {Interaction}: {Extending} {Boundaries}}} \emph{(\bibinfo{series}{{NordiCHI} '10})}. \bibinfo{publisher}{Association for Computing Machinery}, \bibinfo{address}{New York, NY, USA}, \bibinfo{pages}{531--540}.
\newblock
\showISBNx{978-1-60558-934-3}
\urldef\tempurl%
\url{https://doi.org/10.1145/1868914.1868974}
\showDOI{\tempurl}


\bibitem[Wallbaum et~al\mbox{.}(2018)]%
        {wallbaum_supporting_2018}
\bibfield{author}{\bibinfo{person}{Torben Wallbaum}, \bibinfo{person}{Andrii Matviienko}, \bibinfo{person}{Swamy Ananthanarayan}, \bibinfo{person}{Thomas Olsson}, \bibinfo{person}{Wilko Heuten}, {and} \bibinfo{person}{Susanne~C.J. Boll}.} \bibinfo{year}{2018}\natexlab{}.
\newblock \showarticletitle{Supporting {Communication} between {Grandparents} and {Grandchildren} through {Tangible} {Storytelling} {Systems}}. In \bibinfo{booktitle}{\emph{Proceedings of the 2018 {CHI} {Conference} on {Human} {Factors} in {Computing} {Systems}}} \emph{(\bibinfo{series}{{CHI} '18})}. \bibinfo{publisher}{Association for Computing Machinery}, \bibinfo{address}{New York, NY, USA}, \bibinfo{pages}{1--12}.
\newblock
\showISBNx{978-1-4503-5620-6}
\urldef\tempurl%
\url{https://doi.org/10.1145/3173574.3174124}
\showDOI{\tempurl}


\end{thebibliography}

\end{document}